\documentclass[
 aps,
 12pt,
amsmath,amssymb,
nofootinbib,
 reprint,%
]{revtex4-1}

\usepackage{booktabs}
\usepackage{graphicx}
\usepackage{dcolumn}
\usepackage{romannum}
\usepackage{csquotes}
\usepackage{bm}
\usepackage{chemfig}
\usepackage{mathptmx}
\usepackage{wrapfig}
\usepackage{amsfonts,amsmath,amssymb}
\usepackage{amsmath}

\usepackage{subcaption}
\usepackage{wrapfig,float}

\usepackage{xcolor}
\usepackage{tablefootnote}
\usepackage[colorlinks,linkcolor=blue, citecolor = blue, urlcolor = blue]{hyperref}
\usepackage{placeins}
\usepackage{lmodern}
\usepackage{graphicx,epsf}

\usepackage{etoolbox}
\usepackage{orcidlink}
\usepackage{tikz}
\newsavebox\curwrapfig
\makeatletter
\long\def\wrapfiguresafe#1#2#3{%
  \sbox\curwrapfig{#3}%
  \par\penalty-100%
  \begingroup 
    \dimen@\pagegoal \advance\dimen@-\pagetotal 
    \advance\dimen@-\baselineskip 
    \ifdim \ht\curwrapfig>\dimen@ 
      \break%
    \fi%
  \endgroup%
  \begin{wrapfigure}{#1}{#2}%
    \usebox\curwrapfig%
  \end{wrapfigure}%
}

\newcommand{\etal}{\textit{et al.}}

\newcommand{\p}{^{+}}

\makeatletter
\def\@email#1#2{%
 \endgroup
 \patchcmd{\titleblock@produce}
  {\frontmatter@RRAPformat}
  {\frontmatter@RRAPformat{\produce@RRAP{*#1\href{mailto:#2}{#2}}}\frontmatter@RRAPformat}
  {}{}
}%
\makeatother

\begin{document}

\preprint{AIP/123-QED}

\title[]{Electron impact partial ionization cross sections: R-carvone, 2-butanol, imidazole, and 2-nitroimidazole}
\author{Suriyaprasanth Shanmugasundaram  \orcidlink{0000-0002-3939-1446}}
\affiliation{Department of Physics, School of Advanced Sciences,\\
Vellore Institute of Technology, Katpadi,\\
Vellore - 632014, Tamil Nadu, India.} 
\thanks{}

\author{Rounak Agrawal}
\affiliation{School of Computer Science and Engineering,\\
Vellore Institute of Technology, Katpadi,\\
Vellore - 632014, Tamil Nadu, India.}
\author{Dhanoj Gupta$^{*}$\footnote{Corresponding author}\orcidlink{0000-0001-6717-8194} \thanks{Corresponding author:}}
\affiliation{Department of Physics, School of Advanced Sciences,\\
Vellore Institute of Technology, Katpadi,\\
Vellore - 632014, Tamil Nadu, India.} 
\email{dhanoj.gupta@vit.ac.in}


\date{\today}

\begin{abstract}
We {calculate} the electron impact partial and total ionization cross sections of R-carvone $(\mathrm{C_{10}H_{14}O})$, 2-butanol $(\mathrm{C_{4}H_{10}O})$, imidazole $(\mathrm{C_3H_4N_2})$ and 2-nitroimidazole $(\mathrm{C_3H_3N_3O_2})$. We have used the Binary Encounter Bethe (BEB) model to obtain the total electron impact ionization cross sections (TICS). The modified BEB method in combination with mass spectrum data of the molecules is used to calculate the {partial ionization cross section (PICS)} of the cationic fragments dissociating from the parent molecule. Our PICS data for R-carvone and 2-butanol are in good agreement with the experimental data for all the cation fragments along with the TICS data.  For imidazole and 2-nitroimidazole, the estimates of the PICS are reported for the first time in the present study. We have found that both the modified BEB method and the mass spectrum dependence method work effectively to estimate PICS if we have information about the appearance energies and relative abundance data of the target under investigation. 
\end{abstract}

\maketitle

\section{\label{sec:introduction}Introduction:}
Electron impact total ionization cross section (TICS) and partial ionization cross section (PICS) are important to many branches of sciences ranging from plasma physics to astrophysics and biological sciences.\cite{Itikawa2017, KIM2015, Boudaiffa2000} In medical sciences, radiation damage to DNA/RNA is a major concern, where it has been found that secondary electrons play a crucial role in single and double-strand breaks of DNA/RNA.\cite{Boudaiffa2000, Sanche2002} The electron impact cross sections data at low and intermediate energy are important inputs in Monte Carlo track (MCT) analysis, which is used to study damage to living cells using ionizing radiation.\cite{Boudaiffa2000} In recent times, electron collision with biomolecules has geared up and many studies have been conducted both experimentally \cite{rehman2022determination,rehman2022electron} and theoretically \cite{Gupta2014, Gupta2014mol, Gupta2019,vinodkumar2006theoretical, scheer2007total} to provide more data for understanding the mechanism of damages caused due to the electrons. While performing the MCT simulations, only the single channel ionization is taken into account and the channels due to the dissociative ionization processes are not considered in the model, which may reduce the accuracy of the MCT analysis. Therefore, the inclusion of dissociative ionization processes is important and will help us to make the MCT model more accurate. Hence, providing information on the PICS for the cations that are different due to their mass-to-charge ratio originating from their parent molecule is {invaluable} for accurate analysis. \cite{rehman2022electron}
Even with the rapid surge of electron-molecule interaction studies, there are many important molecules for which data are scarce in the literature both experimental and theoretical. In the present study, we have investigated the PICS and TICS of R-carvone, 2-butanol, imidazole, and 2-nitroimidazole. Our motivation has been to provide the theoretical estimates of the PICS for all the cation fragments originating from the parent molecules using the recently developed methods.\cite{graves2022calculated, goswami2022electron} Here we will highlight the importance and applications of each of these targets. 

R-carvone is an organic molecule that belongs to the class of terpenoids and has a wide variety of applications in the food industry, agriculture, and pest control. R-carvone has a minty flavored odor, used in essential oils. A recent study has found this compound can be used in anti-fungal wrapper \cite{boonruang2017antifungal,soleimani2022phenolic} for packaging fruits which increases shelf life. Jones \etal\cite{jones2019dynamical}~performed an experimental and theoretical study on electron (e, 2e) impact ionization dynamics on R-carvone and its enantiomer S-carvone, using three-body distorted wave (3BDW) formalism for calculating the cross sections theoretically. Lopes \etal \cite{lopes2020electron} and Amorim \etal \cite{amorim2021absolute,amorim2021electron} performed a three-part study performed on this molecule. In the first part, Lopes \etal \cite{lopes2020electron} presented the appearance energies for 35 cations and the electron impact mass spectrum in which they observed 103 peaks with the incident electron energy of 70 eV at the mass range of 1 amu till 151 amu and compared their findings with the data available at National Institute of Standards and Technology (NIST) web book \cite{linstrom2001nist} and spectral database for organic compounds (SDBS).\cite{SDBS} Also, they have provided the scheme for the ionic fragmentation pathways.
The successive work of Amorim \etal \cite{amorim2021absolute} presents the PICS for 78 cations from ionization threshold to 100 eV, the PICS of the isotopologue was also measured. The absolute values of PICS for all the cations presented are in the range of 8 eV to 100 eV, only singly ionized fragments were assessed. The last work from Amorim \etal  \cite{amorim2021electron} in this series of articles contains the experimental TICS of the R-carvone compared with literature data along with the theoretically predicted TICS, using the Binary encounter Bethe (BEB) model\cite{amorim2021electron} and the independent atom model (IAM) in conjunction with the screening correction additive rule (SCAR) incorporating the interference effects (I) also, which is colloquially called IAM+SCAR+I method which can be used to calculate the TICS. \cite{blanco2002improvements,blanco2003screening,blanco2003improvements,blanco2016screening,chiari2013total} In all the studies, Lopes \etal~ has used Hiden analytical energy pulse ion counting (EPIC) quadrupole mass spectrometer (QMS - EPIC 300) which can detect mass from 1 amu to 300 amu comprising a detector resolution of 1 amu which was used at residual gas analyzing (RGA) mode.


2-butanol is a colorless organic solvent that also possesses chirality. \cite{iwahashi1999molecular} Recently the compound has become a study of interest for many researchers as it can be a potential replacement for gasoline and can be directly used as a replacement without modifying the engines \cite{lopes2020electronbiofuel} and references therein. Like other alcohols 2-butanol also belongs to the class of biofuels like methanol, ethanol, 1-proponol, and 1-butanol.  Several experimental and theoretical investigations are available for these molecules \cite{pires2018electron,ghosh2018electron,goswami2022electron} and references therein. {Bhavsar} \etal \cite{bhavsar2023dynamics} provided a recent study of electron collision dynamics with n-butanol containing an extensive review of the available literature on n-butanol. But studies on 2-butanol are scarce. Researchers are looking to engineer microbes to produce fuels such as 1-butanol and 2-butanol and have identified an engineered strain of \textit{L. diolivorans} microbe, which can be used to produce 2-butanol up to 10g/Litre from meso-2,3-butanediol through an anaerobic fermentation process.\cite{russmayer2019microbial} Amorim \etal \cite{amorim2022-2butanol-1} experimentally observed the mass spectrum for 51 cations and the appearance energies for 38 cationic fragments of 2-butanol are provided in their work. For their measurements, they used the same QMS EPIC-300 operated on residual gas analyzing (RGA) mode.\cite{nixon2016electron,pires2018electron,lopes2020electron} Lopes \etal ~have performed a comprehensive review on the fragmentation of biofuels which contains all the literature till the year 2019 for the methanol, ethanol, propanol, and butanol molecules.\cite{lopes2020electronbiofuel} Bettega and co-workers performed low-energy electron scattering studies on isomers of butanol including 2-butanol where they presented the differential cross sections (DCS), momentum transfer cross sections (MTCS), and integral cross sections (ICS) for an incident energy range of 1 eV to 50 eV.\cite{bettega2010low} The preceding work of Amorim \etal~provided the experimental PICS for the fragments of 2-butanol along with their experimental TICS and theoretical TICS using the BEB method and the IAM+SCAR+I method.\cite{amorim20232-butanol-2}

Imidazole acts as a base for several biomolecules. Imidazole and 2-nitroimidazole are very good radiosensitizers used in cancer treatment and also in the development of drugs in pharmaceutical chemistry.\cite{meissner2019low} A brief review \cite{shalini2010imidazole,zhang2014comprehensive} of their use provides us with the importance of these targets for various applications. {Meissner \etal} \cite{meissner2019electron} studied the low-energy fragmentation of imidazole and 2-nitroimidazole, and has provided us with the appearance energies and mass spectrum for the cationic fragments. A recent study by Tejas \etal \cite{jani2023theoretical} for imidazole has investigated the dissociative electron attachment (DEA), excitation, and ionization processes by electron impact.

In the present study, we have made use of the appearance energies and mass spectrum data from the literature \cite{lopes2020electron,meissner2019electron,amorim2022-2butanol-1} for all the targets studied here. The structure of the article is as follows: In section \ref{sec:level2}, we have discussed the BEB model for TICS and the computational method to calculate orbital energies required as input in the BEB model. In sections  \ref{sec:level2a}, \ref{sec:2b}, \ref{sec:2c}, various methods to calculate the PICS employing the BEB model using the appearance energies, and the mass spectrum data are discussed. In section \ref{sec:level4}, we have shown the results and their discussions. In Section \ref{sec:level5} we summarize our results and findings. 
\section{\label{sec:level2} BEB Model}
The BEB model \cite{kim1994binary} is one of the most common and widely used methods for predicting the TICS for atoms, molecules, radicals, and ions and is being employed here with some modifications to calculate the PICS. The BEB method calculates the electron impact ionization cross section for each orbital and the sum of the ionization cross section for each orbital gives the total ionization cross section  [Eq. \eqref{eq:1}] for a target under consideration. [Eq. \eqref{eq:2}] gives the BEB formula for determining the TICS.
\begin{equation}\label{eq:1}
    \sigma_{TICS}(E) = \sum_{i}^{N} \sigma_{i}(E)
\end{equation}
\begin{widetext}
\begin{equation}\label{eq:2}
 \sigma_i^{BEB}(E)=\frac{S}{(t_i+u_i+1)/ n}\left [ \frac{ Q_{i} \ln t_i}{2}\left ( 1 - \frac{1}{t_i^{2}}\right) + (2-Q_{i}) \left \{ \left ( 1 - \frac{1}{t_{i}} \right )- \frac{\ln t_{i}}{t_{i}+1} \right \} \right ]
\end{equation}
\end{widetext}
The reduced variables $t_{i},u_{i}$ and S are defined,
\begin{equation}\label{eq:3}
S=4\pi a_0^2 N \left (\frac{R}{B}\right )^2,~u_i=\frac{U}{B},~t_i=\frac{E}{B},~Q_i=1
\end{equation}
Here, $n$ is the {scaling factor that is set to unity in the current calculations for all the molecular orbitals present in our targets as they contain only light atoms $(Z \leq 10)$. For neutral molecular targets consisting of heavy atoms $(Z\geq10)$, $n$ is set equal to the principal quantum number for each molecular orbital if the orbital is dominated by atomic orbitals with the principal quantum number $>2$ as judged by a Mulliken population $>50\%$,\cite{Scott2005}} $a_0$ is the Bohr radius which equals $0.52918~A^{\circ}$, R is Rydberg constant whose value is $13.6057~\mathrm{eV}$, B is the {binding energy of the electron in the orbital}, U is the orbital kinetic energies, N is the orbital occupation number, E is the incident electron kinetic energy and $Q_{i}$ is differential oscillator strength which is set to unity according to the simplified BEB model. 

\begin{figure*}
     \centering
     \begin{subfigure}[b]{0.24\textwidth}
         \centering
         \includegraphics[width=\textwidth]{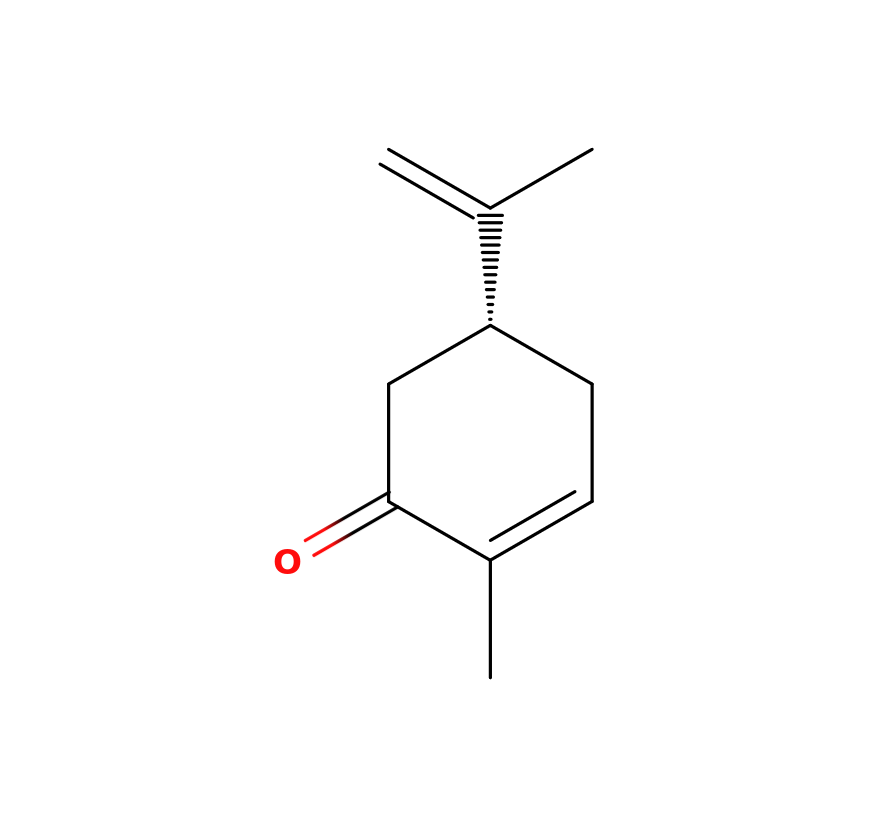}
         \caption{R-carvone $(C_{10}H_{14}O)$}
         \label{fig:R-carvone-Struct}
     \end{subfigure}
     \begin{subfigure}[b]{0.24\textwidth}
         \centering
         \includegraphics[width=\textwidth]{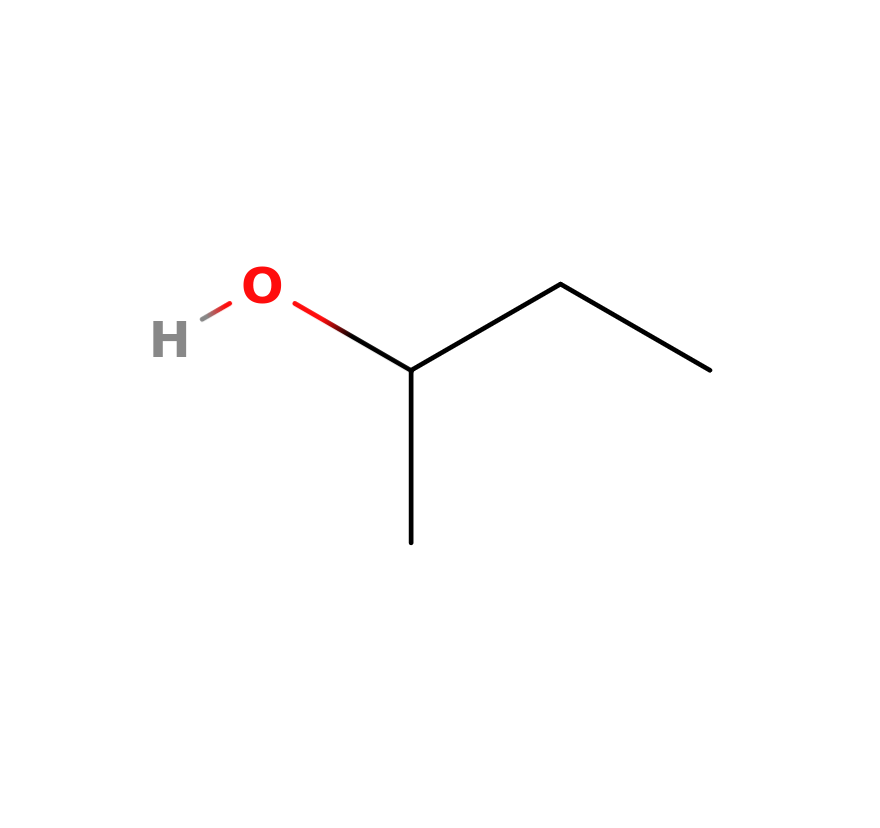}
         \caption{2-butanol $(C_{4}H_{10}O)$}
         \label{fig:2_butanol-Struct}
     \end{subfigure}
     \begin{subfigure}[b]{0.24\textwidth}
         \centering
         \includegraphics[width=\textwidth]{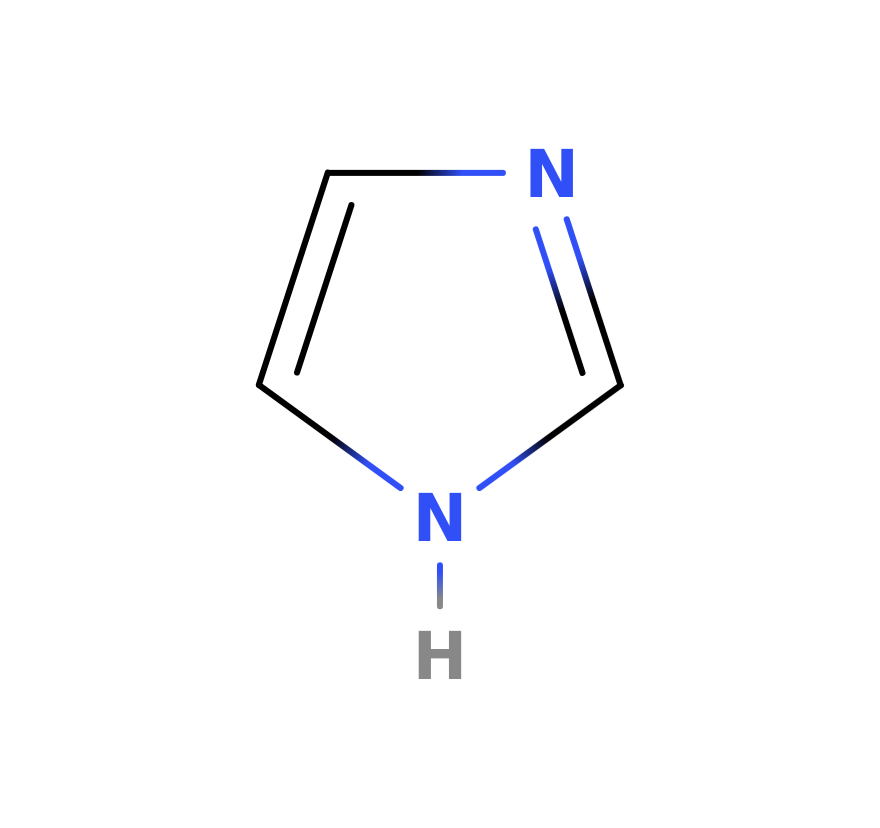}
         \caption{imidazole $(C_3H_4N_2)$}
         \label{fig:IMI-Struct}
     \end{subfigure}
     \begin{subfigure}[b]{0.24\textwidth}
         \centering
         \includegraphics[width=\textwidth]{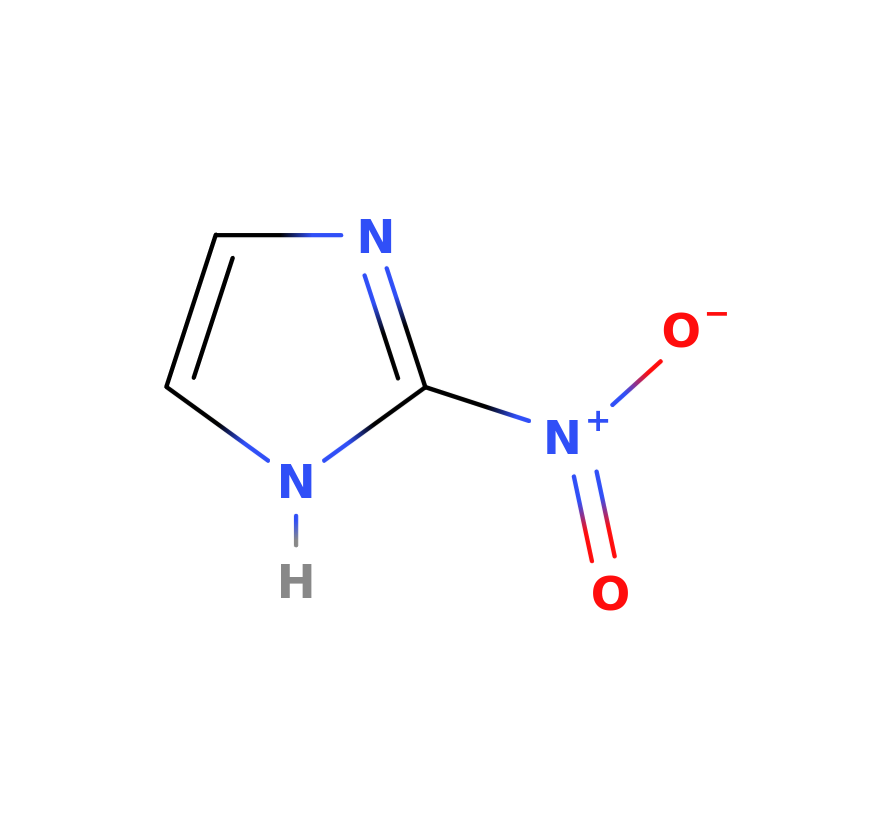}
         \caption{2-nitroimidazole $(C_3H_3N_3O_2)$}
         \label{fig:2NIMI-struct}
     \end{subfigure}
        \caption{Structure of molecular targets, for which the PICS were presented in this manuscript.}
        \label{Structure_fig1}
\end{figure*}

To calculate the orbital parameters that are required as input to the BEB model we adopted the similar methodology that was applied in our recent work, \cite{suriyaprasanth2023electron} we first optimized the target molecule's {geometry with density functional theory (DFT)} using the density functional $\omega B97X-D$ with the aug-cc-PVTZ (aVTZ) basis set, which calculates the energy minima at which the molecule is more stable. After optimization, the orbital binding and kinetic energies are calculated using {Hartree–Fock (HF) method. Normally, the HF method gives slightly higher values of binding energies that may lead to lower values of the TICS compared to experiments as the BEB cross sections are sensitive to the binding energies of the valence shell orbitals. Moreover, TICS calculated using the HF orbital parameters are found to give TICS within 10\% to 15 \% of the experimental uncertainty in the measured TICS.\cite{vukstich2022іонізація} To further improve upon the HF method, many researchers have replaced the binding energies of the valence orbitals with the experimental vertical ionization energies\cite{gupta2017electron} and have sometimes used more accurate methods like outer valence green function (OVGF) propagator method\cite{huber2019total} or coupled cluster singles and doubles (CCSD)\cite{irikura2017partial} methods to compute the orbital energies of valence orbitals and HF method for the core orbitals to improve the BEB TICS agreement with experimental TICS. In our case, we have not made any such corrections to the binding energies of the valence shell orbitals using either the OVGF or the CCSD method. However, in addition to the HF method, we have also used DFT with the $\omega$B97X-D functional and the aVTZ basis set to calculate the orbital parameters and used those parameters to obtain the TICS of the molecular targets under study. The TICS computed using the HF and DFT methods are compared and will be discussed in the results section.} 
All the quantum chemistry calculations are performed using the Gaussian-16 \cite{frisch2016gaussian} package. The structures of the target molecules studied are shown in Figure \ref{Structure_fig1}. The main aspect of this work is to calculate the branching ratios and PICS of the possible fragments forming from the parent molecules which is described in the sections to follow.
\subsection{Branching ratio (BR) and Partial ionization cross section (PICS)}\label{sec:level2a}
 When the molecular target undergoes dissociative ionization (DI) by an electron or positron impact, there will be several reaction channels in which DI takes place. The electron impact mass spectrum (EIMS) is a characterization technique that can provide information about the various cations present, along with their mass-to-charge ratio $(m/z)$, relative abundances, and respective appearance energies. Here, the branching ratio plays an important role, which describes the contribution of each reaction channel to the TICS. There are different methods proposed to calculate the branching ratio (BR), which in turn is used to compute the PICS. Karl Irikura \cite{irikura2017partial} used the EIMS data from the NIST \cite{linstrom2001nist} to calculate BR and have used that to compute the PICS of many organic molecules at a single electron impact energy. Hamilton \etal \cite{hamilton2017calculated} used the experimental EIMS to calculate the BR and hence the PICS, which is discussed in more detail later. Huber \etal \cite{huber2019total} proposed the way to calculate the BR using the appearance energy of the fragments by electron impact. The BR was combined with the TICS from the BEB method to obtain the PICS for various fragments which is discussed in more detail below in Section. \ref{sec:2b}. Hamilton \etal \cite{hamilton2017calculated} proposed a modification in the reduced variable 
 $t_{i} = E/B$, which is given in [Eq. \eqref{eq:4}]. 
\begin{equation}\label{eq:4}
    t'_{i} = \left ( \dfrac{E}{B-{x} } \right )
\end{equation}
Here, ${x}$ can either be the appearance energy (AE) or the dissociation energy {$(\varphi)$} for the specific fragment. The AE and ionization energy (IE) {are not} the same, IE is the minimum energy that is required to remove an electron from a molecule's bound state whereas AE is {the amount of energy needed to be transferred to the neutral molecule (M) to allow for the detection of the fragment ion $m_1^+$ is called the AE of that fragment ion.\cite{gross2006mass} In the present study, we have not calculated the AEs of the cations but have obtained it from the experimental works of Amorim \etal \cite{amorim2022-2butanol-1}, Lopes \etal\cite{lopes2020electron}, Meissner \etal \cite{meissner2019electron} for all the molecular targets.} {However, it is also possible to calculate the dissociation energy $(\varphi)$ of the fragment ions using the methods prescribed by Huber \etal \cite{huber2019total}  and Graves \etal \cite{graves2022calculated}}
After incorporating the modification of [Eq.\eqref{eq:4}], the modified BEB formula given in [Eq.\eqref{eq:5}] can be used to calculate PICS,
\begin{widetext}
\begin{equation} \label{eq:5}
\sigma_i^{mBEB-A}(E)=\frac{S}{(t'_i+u_i+1)/ n}\left [ \frac{ Q_{i} \ln t'_i}{2}\left ( 1 - \frac{1}{t_i^{'2}}\right) + (2-Q_{i}) \left \{ \left ( 1 - \frac{1}{t'_{i}} \right )- \frac{\ln t'_{i}}{t'_{i}+1} \right \} \right ]
\end{equation}
Using [Eq.\eqref{eq:5}] the PICS of each fragment can be calculated as,
\begin{equation}\label{eq:6}
    \sigma^{PICS} (E) = \Gamma_i(E_r)\times \sigma_i^{mBEB-A}(E)
\end{equation}
\end{widetext}
Here, $\Gamma_i(E_r)$ is the experimental BR obtained from the EIMS which is performed at the reference incident electron energy $(E_r)$ of 70 eV or 100 eV.
The branching ratio of the fragment is defined as,
\begin{equation}\label{eq:7}
    {\Gamma_i(E_{r}) = \dfrac{R(E_{r})}{T(E_{r})}}
\end{equation}
$R(E_r)$ denotes the relative intensity of the particular fragment and $T(E_r)$ is the total ion intensity which is the sum of all the relative intensity of the fragments that are detected. In case of unavailability of EIMS data for a molecule of interest, performing quantum chemical mass spectrometry (QCxMS) \cite{grimme2013towards} can provide information on the fragments along with their relative abundances and fragmentation pathways. The project QCxMS is very well tested \cite{koopman2019calculation} and is actively under development for more accurate prediction of the EIMS theoretically.  

On the other hand Baluja \etal \cite{goswami2021electron} presented a similar form of the [Eq.\eqref{eq:5}], where they also call it a modified BEB method, we address their model with the superscript B making it $\sigma^{mBEB-B}$. 
The key assumption in the mBEB-B model is, that the cation that is dissociated from the neutral parent molecule can be distinguished by their ${\varphi}$, which is different from IE that corresponds to the neutral molecule. The vertical IE is the energy of the highest occupied molecular orbital (HOMO). To make the BEB model feasible across cations, a consistent alteration is made in the values of occupied orbitals binding energies. The changes are accomplished by adding the difference $(\delta)$ between IE of the neutral molecule and the ${\varphi}$ of the cation $\delta = (IE-{\varphi})$ to all the values in orbital binding energy (B) of the neutral molecule ($B' = B + \delta$). As a result, the HOMO of the particular cation becomes its appearance energy. 
\begin{widetext}
\begin{equation} \label{eq:8}
\sigma_i^{mBEB-B}(E)=\frac{S^{*}}{(t^{*}_i+u^{*}_i+1)/ n}\left [ \frac{ Q_{i} \ln t^{*}_i}{2}\left ( 1 - \frac{1}{t_i^{*2}}\right) + (2-Q_{i}) \left \{ \left ( 1 - \frac{1}{t^{*}_{i}} \right )- \frac{\ln t^{*}_{i}}{t^{*}_{i}+1} \right \} \right ]
\end{equation}
The new reduced variables are,
\begin{equation}\label{eq:9}
    S^{*}=4\pi a_0^2 N \left (\frac{R}{B'}\right )^2,~t^{*}_{i} = \left ( \dfrac{E}{B'} \right ),~u^{*}_i=\frac{U}{B'},~Q_i=1
\end{equation}

The PICS is calculated using a similar approach used in Eq. \eqref{eq:6}
\begin{equation}\label{eq:10}
    \sigma^{PICS} (E) = \Upsilon_i (E_r) \times \sigma_i^{mBEB-B}(E)
\end{equation}
\end{widetext}

{The PICS calculated from the mBEB-B model [Eq.\eqref{eq:8}] are scaled using the factor $\Upsilon_i(E_r)$ as given in [Eq. \eqref{eq:10}]. The scaling factor $\Upsilon_i(E_r)$ is obtained from the ratio of experimental BR to that of the theoretical BR. The experimental BR was obtained from [Eq. \eqref{eq:7}]. The theoretical BR for the fragment is calculated using the ratio of the PICS of the cation and the TICS of the parent molecule at the energy for which experimental BR was obtained as given below in [Eq. \eqref{eq:11}]}
    \begin{equation}\label{eq:11}
        \Gamma_i^{Theo}(E_r)  = \frac{\sigma_i^{mBEB-B}(E_r) }{\sigma_i^{BEB}(E_r)}
    \end{equation}
There are other ways to calculate the theoretical BR without the need for PICS, which was introduced by Huber \etal \cite{huber2019total} which will be explained in the later part of this article.
The scaling factor $\Upsilon_i (E_r)$ is obtained by
\begin{equation}\label{eq:12}
\Upsilon_i (E_r)  = \frac{\Gamma_i(E_r) }{\Gamma_i^{Theo}(E_r)}
\end{equation}
This scaling factor implemented by Baluja \etal \cite{goswami2021electron} and the structural factor incorporated by Huber \etal \cite{huber2019total} and {Graves \etal} \cite{graves2022calculated} are the same, details about structural factor are discussed in the next section.

\subsection{Huber's method}\label{sec:2b}
To {calculate} the BR of the fragments we require the information about the dissociation energies $({\varphi})$. $b_i$ is the branching factor, which {is not} the same as BR.
\begin{equation}\label{eq:13}
b_i = \left ( \dfrac{1}{{\varphi}} \right )^{\alpha}~ \mathrm{if}~E\geq {\varphi}
\end{equation}
$\alpha$ is a parameter that is used to make the $b_i$ slowly reach the asymptotic region and if the incident energy is less than the ${\varphi}$ then $b_i$ is 0 as the dissociation never commences. In our case, we keep it as 3 as proposed by Huber \etal \cite{huber2019total}
The theoretical BR calculated by Huber's method is given below and is denoted by the superscript H.
\begin{equation}\label{eq:14}
    \Gamma_i^{H} (E_r)=  \begin{cases}
\dfrac{b_i}{\sum^n_i b_i} & \text{ if } E\geq {\varphi} \\
 0 & \text{ if } \mathrm{otherwise}
\end{cases}
\end{equation}
It is provided that the additional structural factor $(\chi_i)$ paves the way for other processes within the threshold region.
\begin{equation}\label{eq:15}
    \Gamma_i(E_r) = \Gamma_i^{H}(E_r)\chi_i(E_r)
\end{equation}
$\Gamma_i^{H}$ is the actual BR, and $\chi_i$ is determined from the experimental BR $(\Gamma_i)$ by taking the ratio of experimental BR to that of theoretical BR calculated using Huber's method at the reference energy,
\begin{equation}\label{eq:16}
    \chi_i(E_r) = \dfrac{\Gamma_i(E_r)}{\Gamma_i^H(E_r)}
\end{equation}
The way this structural factor is calculated is the same as the scaling factor $(\Upsilon_i)$ which was discussed earlier. It is shown that for simple dissociation of $C-H$ bonds, the $\chi_i$ is taken as, $\chi_i\leq1$, and for more complex dissociation it is suggested to be $\chi_i >1$.\cite{huber2019total} It was considered $\chi_i=1$ by {Graves \etal}\cite{graves2022calculated} for the ease of calculation of branching ratios and this could be explored further in the future.
\subsection{Mass spectrum dependence (MSD) method} \label{sec:2c}
In all the above methods the BR is calculated at a single energy, this method was proposed to make the branching ratio dependent on the incident energy of the electron. The BR calculated using this method will be denoted with the superscript (MSD), continuous data for all energy ranges.
\begin{equation}\label{eq:17}
   \Gamma_i^{MSD} (E) =  \begin{cases}
0 & \text{ if } E<{\varphi} \\ 
 \Gamma_i(E_r)\left [ 1 - \left ( \dfrac{{\varphi}}{E} \right )^{\nu} \right ]& \text{if} ~E\geq {\varphi}
\end{cases}
\end{equation}
Here, we would also need a BR at the reference energy, which can either be theoretical BR or experimental BR. Suppose, there are no experimental EIMS data present for the target then we can use the theoretical BR calculated from Huber's method, referred to as the hybrid method by {Graves \etal \cite{graves2022calculated} This equation is only true if we assume that the known branching ratios represent the asymptotic values as the incident energy tends towards infinity. 
In simple words, this indicates that if we utilize the known branching ratios at a specific finite reference energy, they do not accurately replicate the branching ratios at that reference energy. Instead, they result in an error of approximately 15\% for the primary fragments. The error will increase if the reference energy is not sufficiently high, where all the channels are not open. Therefore to make the branching ratios stable, all the channels should be open and the reference energy should be sufficiently high enough for the [Eq. \eqref{eq:17}] to be valid. \cite{graves2022calculated}
}The control parameter $\nu$ was taken as $1.5\pm0.2$ by Janev \etal \cite{janev2004collision} it was taken as 1.5 in this work as an average value. As we have discussed several methods to calculate the branching ratios, we should also remember that the sum of all the branching ratios of the fragments should sum up to give unity $\sum_i \Gamma_i = 1$, regardless of the approaches used. However, this condition may not be satisfied if we do not have information on all the fragmentation or dissociation pathways for a target molecule.

{Several dissociation pathways for the formation of positive ions exist due to the process of chemical ionization. One such pathway is the proton transfer to the analyte molecule (M). The thermodynamic properties of the molecules containing high proton affinity and low electron affinity tend to have fragmentation pathways containing protonated cations $[M+H]^+$ or deprotonated cations $[M-H]^+$. Auto-protonation is a special case of the chemical ionization process called the self-chemical ionization. Here the proton donors are formed due to the ionized species of the M itself which is an unnecessary phenomenon in EIMS. The production of $[M+H]^+$ ions due to auto-protonation is promoted by varying the physical parameters such as the temperature and pressure in the experiment. Adding to this, if the analyte molecule M holds acidic hydrogens or has a volatile character likewise favors the production of the $[M+H]^+$ cations.\cite{gross2006mass} This inadvertently affects the mass spectrum and the ion abundance curves. Such circumstances can result in some inaccuracies in the determination of the partial ionization cross sections in the MSD and m-BEB methods as they are dependent on the relative abundances and AEs of the fragments.}

\section{\label{sec:level4}Results and Discussions}
We have {calculated} the PICS for all the molecular targets which we have shown in Figure \ref{Structure_fig1}, using the mBEB-B method and the MSD method. {We have also presented a table containing the IEs calculated using different methods for reference in Table \ref{tab:comparision_homo}}. Here, we did not employ Huber's method \cite{huber2019total} and the mBEB-A \cite{hamilton2017calculated} method as the magnitude of their cross sections was not as accurate as the MSD method \cite{graves2022calculated} or the mBEB-B method \cite{goswami2021electron} as compared to experimental results. More details can be found in the upcoming subsections.
\begin{table}
    \centering
    \caption{The table contains the Highest Occupied Molecular Oribtial (HOMO) energies calculated with the various methods using the aVTZ basis set.}
    \begin{ruledtabular}
\begin{tabular}{ccc}
Molecule&\multicolumn{2}{c}{Method} \\
&DFT (eV)&HF (eV)\\
\hline
R-carvone& 8.89878 & 9.71046  \\
2-butanol&  9.72521  & 11.79096 \\
imidazole& 8.56076  & 8.79099  \\
2-nitroimidazole& 9.56614  & 9.89030  \\
\end{tabular}
\end{ruledtabular}
    \label{tab:comparision_homo}
\end{table}

\subsection{R-Carvone}
In the series of articles published by  Lopes \etal \cite{lopes2020electron}  and 
Amorim \etal  \cite{amorim2021absolute,amorim2021electron} have extensively studied the fragmentation of R-carvone by electron impact and provided the absolute PICS and TICS along with the mass spectrum and appearance energies and the Wannier exponents for 35 cations. We have calculated the theoretical PICS for these fragments using the MSD method and mBEB-B method based on the data presented by Lopes \etal \cite{lopes2020electron}  and Amorim \etal  \cite{amorim2021absolute,amorim2021electron}
The most stable cations were found to have masses of 39 amu $C_3H_3\p$, 54 amu $C_4H_6\p$, and 82 amu $C_5H_6O\p$, as their relative abundances contribute 40\% to the total ion intensity. Owing to higher contribution in the mass spectrum, their PICS also tend to have higher magnitudes than other cations which can be seen from Table \ref{tab:r-carvone}. In the MSD method the BR is scaled for the incident energy whereas in the mBEB-B method, the binding energies of the molecular orbitals are scaled to the appearance energy of the cations. Although they seem different from one another, but work the same when it comes to calculating the PICS. In Table \ref{tab:r-carvone}, we have shown all the important data that we have obtained while performing our calculations. 
It consists of the BR obtained using Huber's method $(\Gamma_i^{H})$ [Eq. \eqref{eq:14}], theoretical BR $(\Gamma_i^{Theo})$ which is calculated while performing the mBEB-B method [Eq. \eqref{eq:11}], the experimental BR $(\Gamma_i)$ which is obtained from the EIMS data presented by Lopes \etal \cite{lopes2020electron} [Eq. \eqref{eq:7}], the structural factor $(\chi_i)$ [Eq. \eqref{eq:16}], the scaling factor $(\Upsilon_{i})$ [Eq. \eqref{eq:12}] and the PICS max from the MSD method and the mBEB-B method compared with the experimental PICS max.\cite{amorim2021absolute} The mBEB-B PICS presented in the table are scaled using the $(\Upsilon_{i})$. We have also calculated the mean squared error (MSE) for the PICS, assuming that the experimental values are true values and the PICS we obtained as predicted values. The MSE for MSD-PICS was  0.1327 and the MSE for mBEB-B was 0.1038 giving almost similar PICS. 
\begin{figure}
    \centering
    \includegraphics[width=0.45\textwidth]{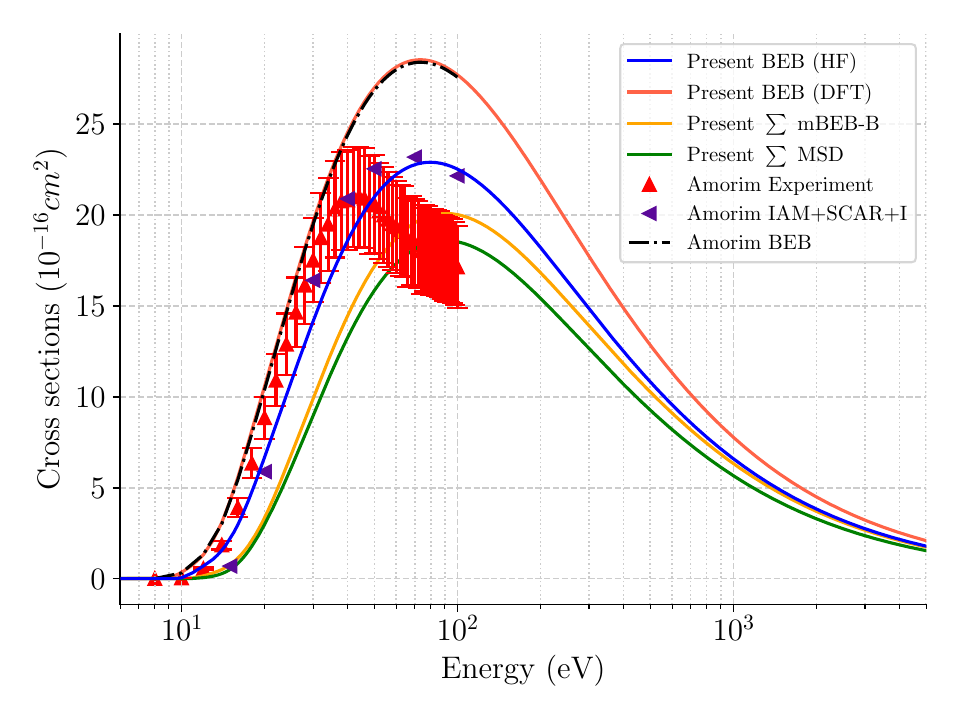}
    \caption{Comparison of TICS of R-carvone with the literature. The blue line represents our present TICS calculated using the BEB method {for the orbital parameters obtained using the HF method, the peach-colored line represents the present BEB data for the orbital parameters obtained using DFT method,} the orange line represents the sum of all the PICS calculated for the various cations using the mBEB-B method, and the green line represents the PICS calculated using the MSD method. The red upright triangle represents the TICS presented by Amorim \etal \cite{amorim2021electron}, the purple left triangle represents the TICS which was calculated using IAM+SCAR+I by Amorim \etal \cite{amorim2021electron} and the black dotted lines represents the TICS calculated using the BEB model by Amorim \etal \cite{amorim2021electron}}
    \label{fig3:tics-r-carvone}
\end{figure}
{In Figure \ref{fig2:pics-r-carvone}, It is seen that for most of the fragments, the PICS calculated using both methods, in general, shows good agreement with the experimental PICS. Along with the PICS of different cationic fragments, the PICS are also calculated for the two isomers of $C_2H_5\p/CHO\p$ and $C_2HO\p/C_3H_5\p$ cationic fragments detected in the mass spectrum having distinct AEs. For a few cations like $O\p,~C_2H_2\p,~C_3H_4\p,~C_4H_5\p,~$ the PICS calculated using the MSD shows a very good agreement with the experimental data of Amorim \etal \cite{amorim2021absolute}, whereas for several other cations both the MSD method and the mBEB-B method predicts the PICS within the experimental uncertainty as shown Figure \ref{fig2:pics-r-carvone}. However, for a few cations $CH_2\p$, $C_3H_3\p$, $C_7H_6O\p$ and $C_3H_5\p$ slight disagreement of the present data is seen with the experimental results.}
 Figure \ref{fig3:tics-r-carvone}, shows the TICS obtained using the BEB model, MSD, and mBEB-B methods which is the sum of the PICS of all the fragments. Our results are compared with the existing experimental and theoretical data,\cite{amorim2021absolute, amorim2021electron} {and the present results using the mBEB-B and MSD method are found to have a poor agreement and underestimate the experimental data at around of 30 $\sim$ 40 eV.} {In the series of articles by Lopes \etal\cite{lopes2020electron}~and Amorim \etal~\cite{amorim2021absolute,amorim2021electron} they have presented the experimental PICS at 100 eV for 78 cationic fragments but have only provided the AEs for 35 prominent cations. We believe if we could calculate the PICS for other remaining fragments and sum it then it might give a better comparison with the experimental TICS and BEB TICS  at the peak and the shift in the peak may also be improved. Interestingly our TICS calculated using the DFT ($\omega B97X-D$/aVTZ) method shows excellent agreement with the BEB TICS calculated by Amorim \etal \cite{amorim2021electron} calculated using the combination of DFT (B3LYP/aug-cc-pVDZ), OVGF, and experimental ionization energies. However, both the TICS calculated overestimate all other data as can be seen in Figure \ref{fig3:tics-r-carvone}. The TICS calculated from the HF orbital parameters shows good agreement with the experimental data and the IAM+SCAR+I method.}    

\begin{figure*}
    \centering
    \begin{tikzpicture}
    \node at (0,0) {\includegraphics[width=\textwidth]{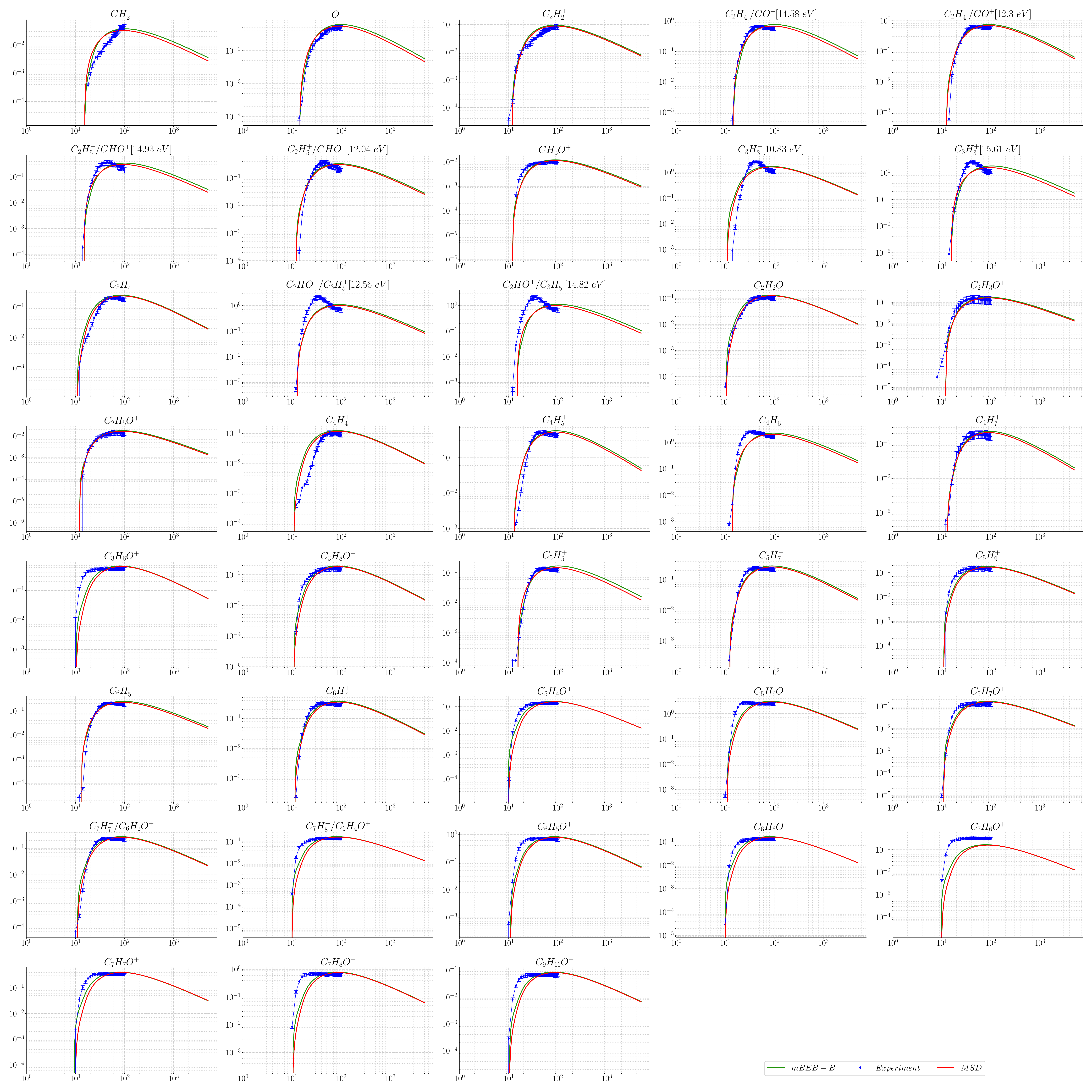}};

    \node[below, yshift=-0.5ex] at (current bounding box.south) {Energy (eV)};

    \node[left, rotate=90, xshift=9.5ex] at (current bounding box.west) {PICS $(10^{-16}cm^{2})$};
  \end{tikzpicture}
    \caption{Comparison of PICS calculated for the cations of R-carvone. The X-axis represents the incident kinetic energy range in (eV), and the Y-axis represents the PICS in $(10^{-16}cm^{2})$. The green line represents the PICS calculated using the mBEB-B method, the red line represents the PICS calculated using the MSD method, and the blue Rhombus represents the experimental PICS presented by Amorim et al. \cite{amorim2021absolute}, along with their error bars for each energy value.}
    \label{fig2:pics-r-carvone}
\end{figure*}
\begin{table*}
    \centering
    \begin{ruledtabular}
        \caption{Absolute values of branching ratio $(\Gamma)$ structural factor $(\chi_i)$, scaling factor $(\Upsilon)$ and maximum PICS calculated using MSD and mBEB-B method compared with experimental PICS \cite{amorim2021absolute}  for R-carvone}
    \begin{tabular}{lllllllllll}  
 $m/z$  & Cations\footnote{The forward slash denotes "or" added for the cations with same $m/z$ values} & ${AE}$ (eV) \cite{lopes2020electron} & $\Gamma_i^H$ & $\Gamma_i^{Theo}$&$\Gamma_i$\cite{lopes2020electron}   &$\chi_i$& $\Upsilon_i$ &$\sigma_{max}^{MSD}$ &$\sigma_{max}^{mBEB-B}$ &$\sigma_{max}^{exp}$\cite{amorim2021absolute}\\  
\hline
 14&$CH_{2}$&15.34&0.033333&0.540558&0.001539 &0.046171&0.002847 &0.03263 &0.037382 &0.04474\\
 16&O&14.16&0.030769&0.608827&0.002601 &0.084534&0.004271 &0.055578 &0.062124 &0.05242\\
 26&$C_{2}H_{2}$&11.67&0.025358&0.794778&0.004087 &0.161172&0.005142 &0.08878 &0.094934 &0.08685\\
 28&$C_2H_4/CO$&12.3&0.026727&0.741232&0.032002 &1.197369&0.043174 &0.692441 &0.748003 &0.64644\\
 28&$C_{2}H_{4}/CO$&14.58&0.031681&0.583319&0.032002 &1.010126&0.054862 &0.682016 &0.768848 &0.64644\\
 29&$C_{2}H_{5}/CHO$&12.04&0.026162&0.762717&0.014356 &0.548735&0.018822 &0.311144 &0.334674 &0.37457\\
 29&$C_{2}H_{5}/CHO$&14.93&0.032442&0.563107&0.014356 &0.442517&0.025494 &0.305212 &0.346601 &0.37457\\
 31&$CH_{3}O$&11.97&0.026010&0.768646&0.000531 &0.020415&0.000690 &0.011508 &0.012364 &0.01021\\
 39&$C_{3}H_{3}$&10.83&0.023533&0.874773&0.074061 &3.147148&0.084663 &1.617523 &1.708448 &2.61851\\
 39&$C_{3}H_{3}$&15.61&0.033919&0.526334&0.074061 &2.183447&0.140712 &1.567225 &1.806174 &2.61851\\
 40&$C_{3}H_{4}$&10.63&0.023098&0.895457&0.010667 &0.461812&0.011913 &0.233269 &0.245712 &0.2041\\
 41&$C_{2}HO/C_{3}H_{5}$&12.56&0.027292&0.720556&0.046743 &1.712707&0.06487 &1.009673 &1.095517 &2.15466\\
 41&$C_2HO/C_{3}H_{5}$&14.82&0.032203&0.569361&0.046743 &1.451525&0.082097 &0.994528 &1.126763 &2.15466\\
 42&$C_{2}H_{2}O$&10.32&0.022425&0.928897&0.005758 &0.256772&0.006199 &0.126158 &0.132352 &0.11159\\
 43&$C_{2}H_{3}O$&11.99&0.026053&0.766946&0.007496 &0.287718&0.009774 &0.162525 &0.174674 &0.15451\\
 45&$C_{2}H_{5}O$&11.99&0.026053&0.766946&0.000756 &0.029017&0.000986 &0.016396 &0.017622 &0.01444\\
 52&$C_{4}H_{4}$&10.68&0.023207&0.890223&0.005387 &0.232130&0.006051 &0.117758 &0.124124 &0.10637\\
 53&$C_{4}H_{5}$&12.52&0.027205&0.723686&0.024386 &0.896380&0.033697 &0.526899 &0.571304 &0.50587\\
 54&$C_{4}H_{6}$&13.81&0.030008&0.631203&0.093897 &3.129061&0.148758 &2.011533 &2.233011 &2.43499\\
 55&$C_{4}H_{7}$&12.67&0.027531&0.712042&0.009911 &0.359995&0.013919 &0.213932 &0.232561 &0.19521\\
 58&$C_{3}H_{6}O$&10.01&0.021751&0.964127&0.029481 &1.355390&0.030578 &0.647126 &0.676302 &0.56612\\
 60&$C_{3}H_{8}O$&10.62&0.023076&0.896510&0.000836 &0.036227&0.000932 &0.01828 &0.019252 &0.01595\\
 65&$C_{5}H_{5}$&15.31&0.033267&0.542169&0.006979 &0.209785&0.012872 &0.147989 &0.169430 &0.1409\\
 67&$C_{5}H_{7}$&12.16&0.026423&0.867692&0.012153 &0.459945&0.016146 &0.263205 &0.283664 &0.24021\\
 69&$C_{5}H_{9}$&10.94&0.023772&0.863681&0.007894 &0.332075&0.00914 &0.172297 &0.182264 &0.15821\\
 77&$C_{6}H_{5}$&13.12&0.028509&0.678578&0.010561 &0.370448&0.015564 &0.227293 &0.249071 &0.2091\\
 79&$C_{6}H_{7}$&11.13&0.024185&0.844976&0.016200&0.669847&0.019172 &0.353152 &0.374592 &0.32259\\
 80&$C_{5}H_{4}O$&9.63&0.020925&1.009962&0.007271 &0.347475&0.007199 &0.159963 &0.16644 &0.14366\\
 82&$C_{5}H_{6}O$&10.85&0.023576&0.872742&0.132679 &5.627668&0.152025 &2.897386 &3.061113 &2.71195\\
 83&$C_{5}H_{7}O$&10.9&0.023685&0.867692&0.007231 &0.305301&0.008334 &0.157858 &0.166896 &0.13467\\
 91&$C_{7}H_{7}/C_6H_3O$&10.83&0.023533&0.874773&0.012047 &0.511925&0.013772 &0.263116 &0.277906 &0.23679\\
 92&$C_{7}H_{8}/C_6H_4O$&9.87&0.021447&0.980661&0.00743 &0.346440&0.007577 &0.163231 &0.170305 &0.14487\\
 93&$C_{6}H_{5}O$&10.83&0.023533&0.874773&0.036261 &1.540875&0.041452 &0.791954 &0.836472 &0.73009\\
 94&$C_{6}H_{6}O$&9.72&0.021121&0.998829&0.007258 &0.343642&0.007266 &0.159585 &0.166215 &0.13833\\
 106&$C_{7}H_{6}O$&9.56&0.020773&1.018745&0.007258 &0.349394&0.007124 &0.159737 &0.166078 &0.3352\\
 107&$C_{7}H_{7}O$&9.35&0.020317&1.045762&0.01815 &0.893349&0.017356 &0.399984 &0.414931 &0.35451\\
 108&$C_{7}H_{8}O$&10.2&0.022164&0.942315&0.034881 &1.573783&0.037017 &0.764773 &0.801114 &0.69171\\
 135&$C_{9}H_{11}O$&10.27&0.022316&0.934455&0.003728 &0.167056&0.003990 &0.081708 &0.085665 &0.07175
 \end{tabular}
\label{tab:r-carvone}
\end{ruledtabular}
\end{table*}

\subsection{2-butanol}
The PICS that we have {determined theoretically} for the fragments of 2-butanol are based on the EIMS data presented by Amorim \etal \cite{amorim2022-2butanol-1} {in which they have presented the AEs and the PICS for 38 cations.} Pires \etal \cite{pires2018electron} and Gosh \etal \cite{ghosh2018electron} provided the EIMS, AEs, and the experimental PICS for the fragments of 1-butanol. Goswami \etal \cite{goswami2022electron} have {calculate}d the PICS using the mBEB-B method [Eq. \eqref{eq:8}], which has good agreement with the existing experimental PICS presented by Pires \etal \cite{pires2018electron} and the experimental TICS presented by Gosh \etal \cite{ghosh2018electron} Owing to a good comparison of the mBEB-B model with the literature data in 1-butanol, we proceed to calculate the PICS theoretically applying both the mBEB-B model and the MSD model that requires the experimental EIMS data. In Table \ref{tab:2-butanol}, all the relevant data required for calculating the PICS using the mBEB-B and MSD are given similarly to R-carvone. The maximum cross sections are also provided in the table using the two methods and are compared with the experimental maximum cross section. In Figure \ref{fig4:PICS-2butanol}, we have compared our PICS data with the experimental PICS presented by Amorim \etal \cite{amorim20232-butanol-2} in the entire energy range. The MSD method and mBEB-B method for 2-butanol show excellent comparison with the experimental data of Amorim \etal \cite{amorim20232-butanol-2} for most of the cationic fragments. However, for a few fragments such as $CH\p$, $C_2H_2\p$, $C_3H_2\p$, and $C_3H_3\p$, there were slight discrepancies when comparing our data with the experimental PICS which can be seen in Figure \ref{fig4:PICS-2butanol}. The PICS that we calculated are consistent for all cations from their AE threshold to 5000 eV, magnitudes of the PICS of the cations differ concerning their RI and AE. 

\begin{figure}
    \centering
    \includegraphics[width=0.45\textwidth]{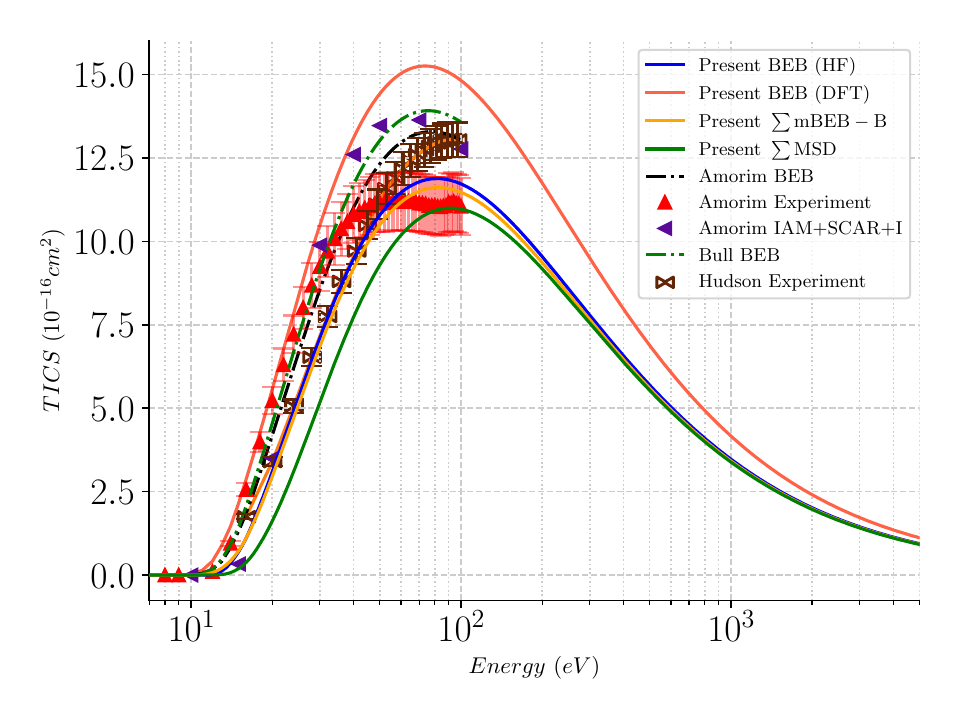}
    \caption{Comparison of TICS of 2-butanol. The blue solid line represents the TICS calculated using the BEB method {for the orbital parameters obtained using the HF method, the peach colored line represents the present BEB data for the orbital parameters obtained using DFT method,} the orange solid line represents the sum of all the PICS calculated by us using the mBEB-B method, the green solid lines represent the sum of all the PICS calculated by us using the MSD method. The back dashed-dotted lines represent the BEB TICS presented by Amorim \etal \cite{amorim20232-butanol-2}, the red upright triangle represents the experimental TICS presented by Amorim \etal \cite{amorim20232-butanol-2} and their error bars are shown in pale pink colors, they have also provided the TICS calculated using the IAM+SCAR+I model which is shown by purple left triangle, the green dashed-dotted line represents the TICS presented by Bull \etal \cite{bull2012absolute} and the brown bowtie represents the experimental TICS presented by Hudson \etal \cite{hudson2003absolute}.}
    \label{fig:2_butanol_TICS}
\end{figure}
Figure \ref{fig:2_butanol_TICS}, shows the comparison of our TICS with the existing data available in the literature for 2-butanol. The present TICS calculated using the BEB method, compares very well with the experimental data and lies within the uncertainty of the experimental TICS. Also the TICS (obtained after summing all the PICS of the fragments) from the mBEB-B and MSD methods compare well with the experiment and are within the range of uncertainty of the experimental TICS. The average experimental uncertainty is $19\%$ and all our present data lies within that limit, {except the one calculated using the DFT $(\mathrm{\omega B97X-D/avTZ})$ method which is higher in magnitude at the peak compared to all other data. Such overestimation of TICS using the DFT $(\mathrm{\omega B97X-D/avTZ})$ is also seen for the fluorocarbon species \cite{gupta2017electron} that may be due to the lower values of the binding energies obtained from the DFT method as the BEB TICS are sensitive to the binding energies of the molecular orbitals.}
\begin{figure*}[h!]
    \centering
      \begin{tikzpicture}
    \node at (0,0) {\includegraphics[width=\textwidth]{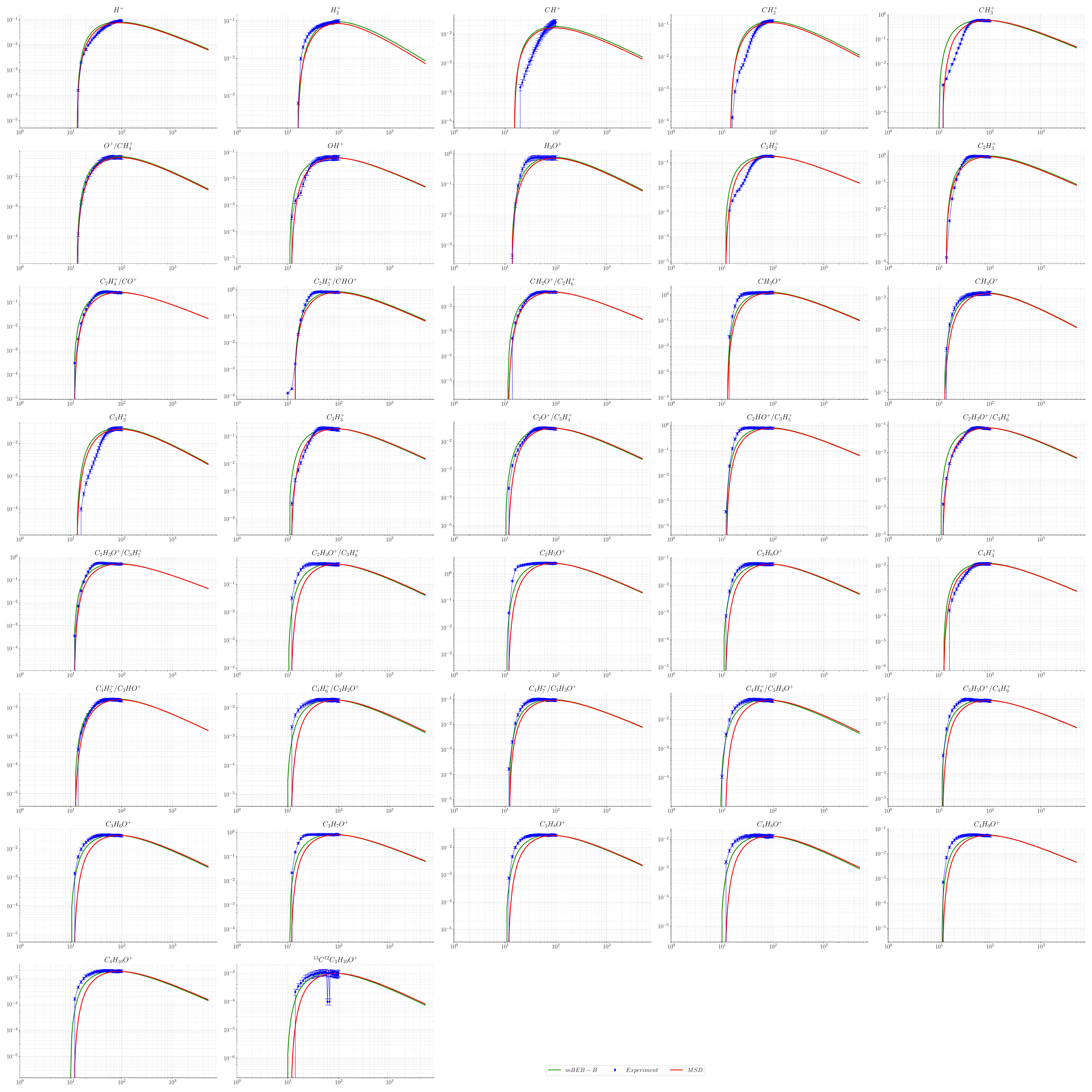}};

    \node[below, yshift=-0.5ex] at (current bounding box.south) {Energy (eV)};

    \node[left, rotate=90, xshift=9.5ex] at (current bounding box.west) {PICS $(10^{-16}cm^{2})$};
  \end{tikzpicture}
    \caption{Comparison of PICS calculated using various methods for 2-butanol. The X-axis represents the incident kinetic energy range in (eV), and the Y-axis represents the PICS in $(10^{-16}cm^{2})$. The green line represents the PICS calculated using the mBEB-B method, the red line represents the PICS calculated using the MSD method, and the blue Rhombus represents the experimental PICS presented by Amorim et al. \cite{amorim20232-butanol-2}, along with their error bars for each energy value.}
    \label{fig4:PICS-2butanol}
\end{figure*}

\begin{table*}[ht!]
    \caption{Absolute values of branching ratio $(\Gamma)$, structural factor $(\chi_i)$, scaling factor $(\Upsilon)$ and maximum PICS calculated using MSD and mBEB-B method compared with experimental PICS \cite{amorim20232-butanol-2}  for 2-butanol.}
\begin{ruledtabular}
    \centering
    \begin{tabular}{llllllllllll}
$m/z$&  Cations\footnote{The forward slash denotes "or" added for the cations with same $m/z$ values} &  ${AE}$ (eV) \cite{amorim2022-2butanol-1} &$\Gamma_i^{H}$&   $\Gamma_i^{Theo}$&$\Gamma_i$\cite{amorim2022-2butanol-1}&  $\chi_i$&$\Upsilon_i$ & $\sigma_{max}^{MSD}$ &$\sigma^{mBEB-B}_{max}$&$\sigma_{max}^{Exp}$ \cite{amorim20232-butanol-2}\\ \hline
1&  $H$&  13.45&0.015970&   0.826677&0.029448&  1.843945&0.035623 &  0.328767&0.354294&0.0929\\
2&  $H_{2}$&  15.52&0.010394&   0.662681&0.033981&  3.269164&0.051278 &  0.374218&0.417960&0.0997\\
13&  $CH$&  15.38&0.010681&   0.672343&0.033674&  3.152747&0.050085 &  0.371191&0.413481&0.0254\\
14&  $CH_{2}$&  14.84&0.011890&   0.711384&0.032492&  2.732773&0.045674 &  0.359453&0.396460&0.1330\\
15&  $CH_{3}$&  9.99&0.038974&   1.251373&0.021873&  0.561216&0.017479 &  0.249473&0.257970&0.594\\
16&  $O/CH_4$&  13.3&0.016517&   0.840578&0.02912&  1.763079&0.034643 &  0.325416&0.349881&0.0459\\
17&  $OH$&  10.91&0.029923&   1.113007&0.023887&  0.798291&0.021462 &  0.270959&0.282552&0.0655\\
19&  $H_{3}O$&  13.56&0.015585&   0.816679&0.029689&  1.905022&0.036354 &  0.331219&0.357553&0.772\\
 26& $C_{2}H_{2}$& 11.5&0.025549&  1.035467&0.025179& 0.985502&0.024317 &  0.284586&0.298677&0.188\\
 27& $C_{2}H_{3}$& 13.87&0.014563&  0.789364&0.030368& 2.085312&0.038472 &  0.338110&0.366797&0.964\\
 28& $C_{2}H_{4}/CO$& 11.59&0.024959&  1.024311&0.025376& 1.016714&0.024774 &  0.286654&0.301163&0.279\\
 29& $C_{2}H_{5}/CHO$& 13.62&0.015380&  0.811294&0.029821& 1.939005&0.036757 &  0.332554&0.359333&0.834\\
 30& $CH_2O/C_{2}H_{6}$& 11.46&0.025818&  1.040480&0.025091& 0.971845&0.024115 &  0.283665&0.297576&0.0392\\
 31& $CH_{3}O$& 12.91&0.018059&  0.878236&0.028266& 1.565195&0.032185 &  0.316662&0.338509&1.27\\
 32& $CH_{4}O$& 12.5&0.019895&  0.920350&0.027368& 1.375620&0.029737 &  0.307408&0.326724&0.015\\
 38& $C_{3}H_{2}$& 13.11&0.017245&  0.858645&0.028704& 1.664470&0.033429 &  0.321159&0.344321&0.0289\\
 39& $C_{3}H_{3}$& 10.52&0.033375&  1.168847&0.023033& 0.690118&0.019706 &  0.261885&0.272052&0.198\\
 40& $C_2O/C_{3}H_{4}$& 10.35&0.035047&  1.194444&0.022661& 0.646585&0.018972 &  0.257912&0.267511&0.0317\\
 41& $C_2HO/C_{3}H_{5}$& 12.34&0.020679&  0.937536&0.027018& 1.306544&0.028818 &  0.303780&0.322175&0.801\\
 42& $C_2H_2O/C_{3}H_{6}$& 10.66&0.032078&  1.148352&0.02334& 0.727609&0.020325 &  0.265148&0.275806&0.0805\\
 43& $C_2H_3O/C_{3}H_{7}$& 11.77&0.023831&  1.002498&0.02577& 1.081357&0.025706 &  0.290782&0.306157&0.566\\
 44& $C_2H_4O/C_{3}H_{8}$& 10.35&0.035047&  1.194444&0.022661& 0.646585&0.018972 &  0.257912&0.267511&0.546\\
 45& $C_{2}H_{5}O$& 10.93&0.029759&  1.110246&0.023931& 0.804168&0.021555 &  0.271423&0.283094&2.53\\
 46& $C_{2}H_{6}O$& 10.67&0.031988&  1.146908&0.023362& 0.730346&0.020369 &  0.265381&0.276075&0.0629\\
 51& $C_{4}H_{3}$& 12.33&0.020729&  0.938625&0.026996& 1.302309&0.028761 &  0.303552&0.321891&0.0118\\
 53& $C_{4}H_{5}/C_3HO$& 12.01&0.022431&  0.974409&0.026296& 1.172314&0.026986 &  0.296266&0.312863&0.0198\\
 54& $C_{4}H_{6}/C_3H_2O$& 9.75&0.041924&  1.291528&0.021347& 0.509186&0.016529 &  0.243820&0.251661&0.0192\\
 55& $C_{4}H_{7}/C_3H_3O$& 12.34&0.020679&  0.937536&0.027018& 1.306544&0.028818 &  0.303780&0.322175&0.102\\
 56& $C_{4}H_{8}/C_3H_4O$& 9.45&0.046045&  1.344407&0.020691& 0.449368&0.01539 &  0.236726&0.243829&0.0483\\
 57& $C_3H_5O/C_{4}H_{9}$& 11.38&0.026366&  1.050608&0.024916& 0.944997&0.023716 &  0.281823&0.295374&0.0974\\
 58& $C_{3}H_{6}O$& 10.04&0.038395&  1.243236&0.021982& 0.572524&0.017682 &  0.250648&0.259290&0.0313\\
 59& $C_{3}H_{7}O$& 10.99&0.029274&  1.102023&0.024062& 0.821959&0.021835 &  0.272813&0.284722&0.84\\
 60& $C_{3}H_{8}O$& 10.66&0.032078&  1.148352&0.02334& 0.727609&0.020325 &  0.265148&0.275806&0.0298\\
 72& $C_{4}H_{8}O$& 9.66&0.043107&  1.307068&0.02115& 0.490645&0.016182 &  0.241695&0.249305&0.014\\
 73& $C_{4}H_{9}O$& 11.45&0.025886&  1.041738&0.02507& 0.968492&0.024065 &  0.283435&0.297300&0.0588\\
 74& $C_{4}H_{10}O$& 9.98&0.039092&  1.253010&0.021851& 0.558970&0.017439 &  0.249238&0.257706&0.0196\\
 75& $^{13}C^{12}C_{3}H_{10}O$& 9.79&0.041412&  1.284707&0.021435& 0.517603&0.016685 &  0.244764&0.252710&0.00111\\
    \end{tabular}
    \label{tab:2-butanol}
\end{ruledtabular}
\end{table*}
\subsection{Imidazole (IMI) and 2-nitroimidazole (2NI)} 
\begin{figure*}
    \begin{subfigure}[b]{0.45\textwidth}
    \centering
     \begin{tikzpicture}
    \node at (0,0) {\includegraphics[width=\textwidth]{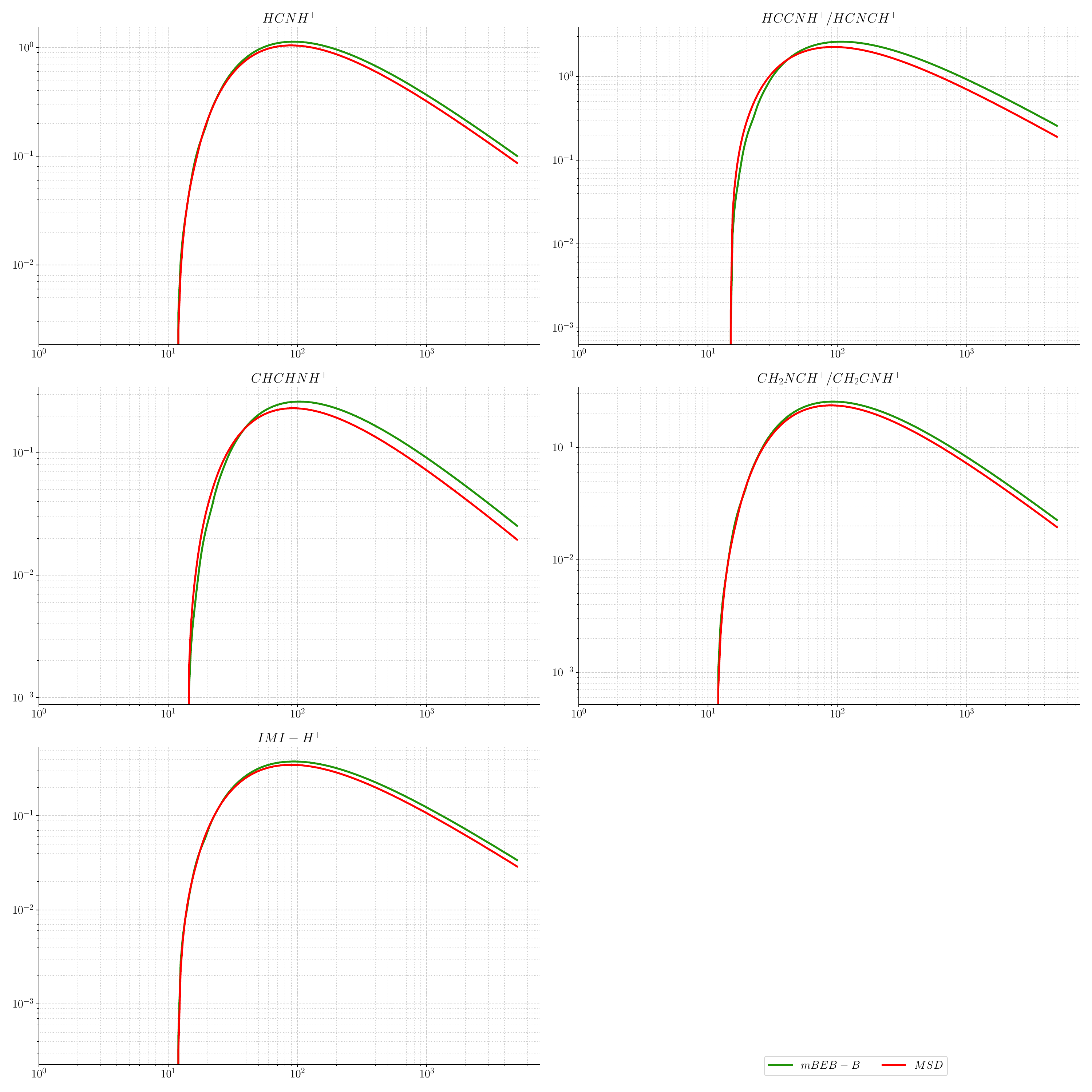}};

    \node[below, yshift=-0.5ex] at (current bounding box.south) {Energy (eV)};

    \node[left, rotate=90, xshift=9.5ex] at (current bounding box.west) {PICS $(10^{-16}cm^{2})$};
  \end{tikzpicture}
    \caption{}
    \label{fig:pics-imidzaole}
\end{subfigure}
\hfill
\begin{subfigure}[b]{0.45\textwidth}
    \centering
    \includegraphics[width=\textwidth]{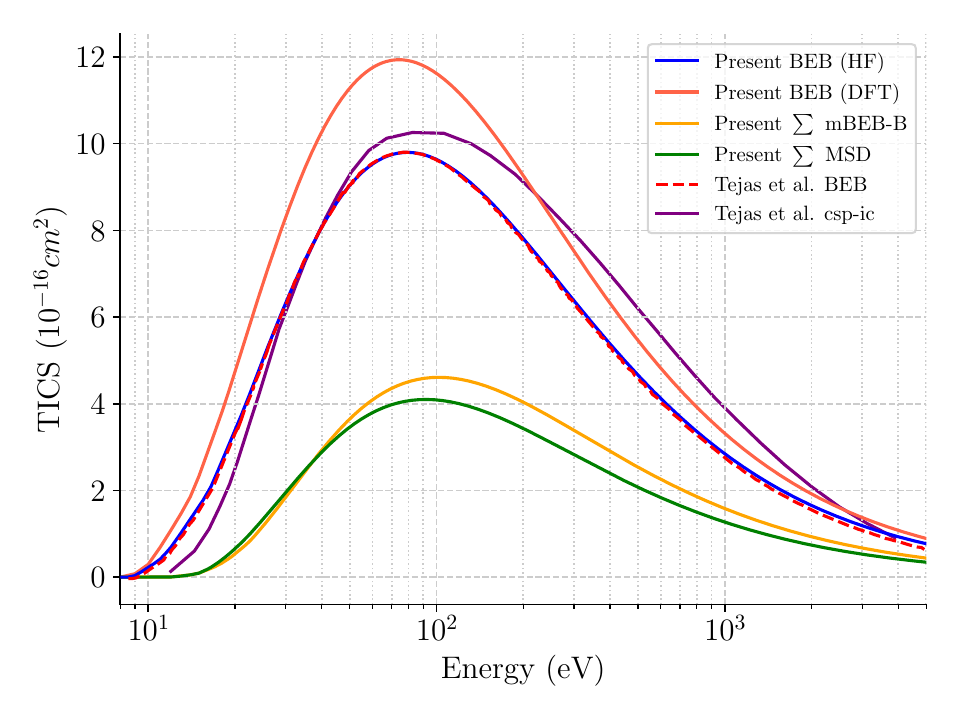}
    \caption{}
    \label{fig:imidazole-tics}
\end{subfigure}
\caption{a) Comparison of PICS calculated using various methods for imidazole. The green line represents the PICS calculated using the mBEB-B method, and the red line represents the PICS calculated using the MSD method, b) Comparison of TICS for imidazole, the blue line indicates the present BEB data {for the orbital parameters obtained using the HF method, the peach colored line represents the present BEB data for the orbital parameters obtained using DFT method,} the orange line indicates the sum of PICS calculated using the mBEB-B method, the green line represents the sum of the PICS calculated using the MSD method, {the black triangles}, and the purple line represents the BEB TICS and the csp-ic TICS data presented by Tejas \etal \cite{jani2023theoretical}.}
\end{figure*}

The PICS that we present here for IMI and 2NI are based on the mass spectrum and AEs presented by {Meissner \etal}\cite{meissner2019electron} In Table \ref{tab:imidazole-table}, and Table \ref{tab:2Nimi_table}, we have presented the BRs calculated using Huber's method [Eq. (\ref{eq:14})], the theoretical BRs [Eq. (\ref{eq:11})], the experimental BRs calculated from the EIMS data [Eq. (\ref{eq:7})] and the structural factor [Eq.(\ref{eq:16})] and the scaling factor [Eq. (\ref{eq:12})] along with the maximum cross sections calculated using the MSD and the mBEB-B methods. The PICS for $HCNH\p$, $HCCNH\p$, $HCNCH\p$ and $CHCHNH\p$ cationic fragments of IMI are shown in Figure \ref{fig:pics-imidzaole}, which were observed in the experiment of {Meissner \etal}\cite{meissner2019electron} $HCNH\p$ is the cation with the lowest m/z having the second highest AE of 11.74 eV, $HCCNH\p$ and $HCNCH\p$ fragments have  m/z value of 40 and same AE, $CH_2NCH\p$ \& $CH_2CNH\p$ also share their AEs and their m/z value are same 41, $CHCHNH\p$ fragment has the same m/z value of 41 but has a different AE and $IMI-H\p$ is the de-hydrogenated IMI which was also observed in the EIMS experiment. 
{It is to be noted that for fragments that have the same m/z values and AEs, the calculated PICS will be the same.}  From Figure \ref{fig:pics-imidzaole}, it can be seen that for almost all the fragments the magnitude of the PICS from the MSD method is slightly smaller than the PICS calculated using the mBEB-B method. Both methods match very well from the ionization threshold to 70 eV after which the cross sections show slight variation for all the fragments.

In Figure \ref{fig:imidazole-tics}, we have compared all the TICS from the existing literature along with our data. Our BEB TICS compares very well with the BEB TICS presented by Tejas \etal \cite{jani2023theoretical}, albeit their TICS calculated using the {complex scattering potential – ionization contribution \cite{Gupta2014} (CSP-ic) method having} a higher magnitude in comparison to BEB data, especially at higher energies after the peak. The magnitudes of the sums of the PICS calculated using both the MSD and the mBEB-B methods are lower than the TICS due to the lack of information on all the fragments and the AEs as our methods are dependent on the AEs and mass spectrum data of the fragments of the parent molecule.

\begin{figure*}[h]
\begin{subfigure}[b]{0.45\textwidth}
    \centering
    \begin{tikzpicture}
    \node at (0,0) {     \includegraphics[width=\textwidth]{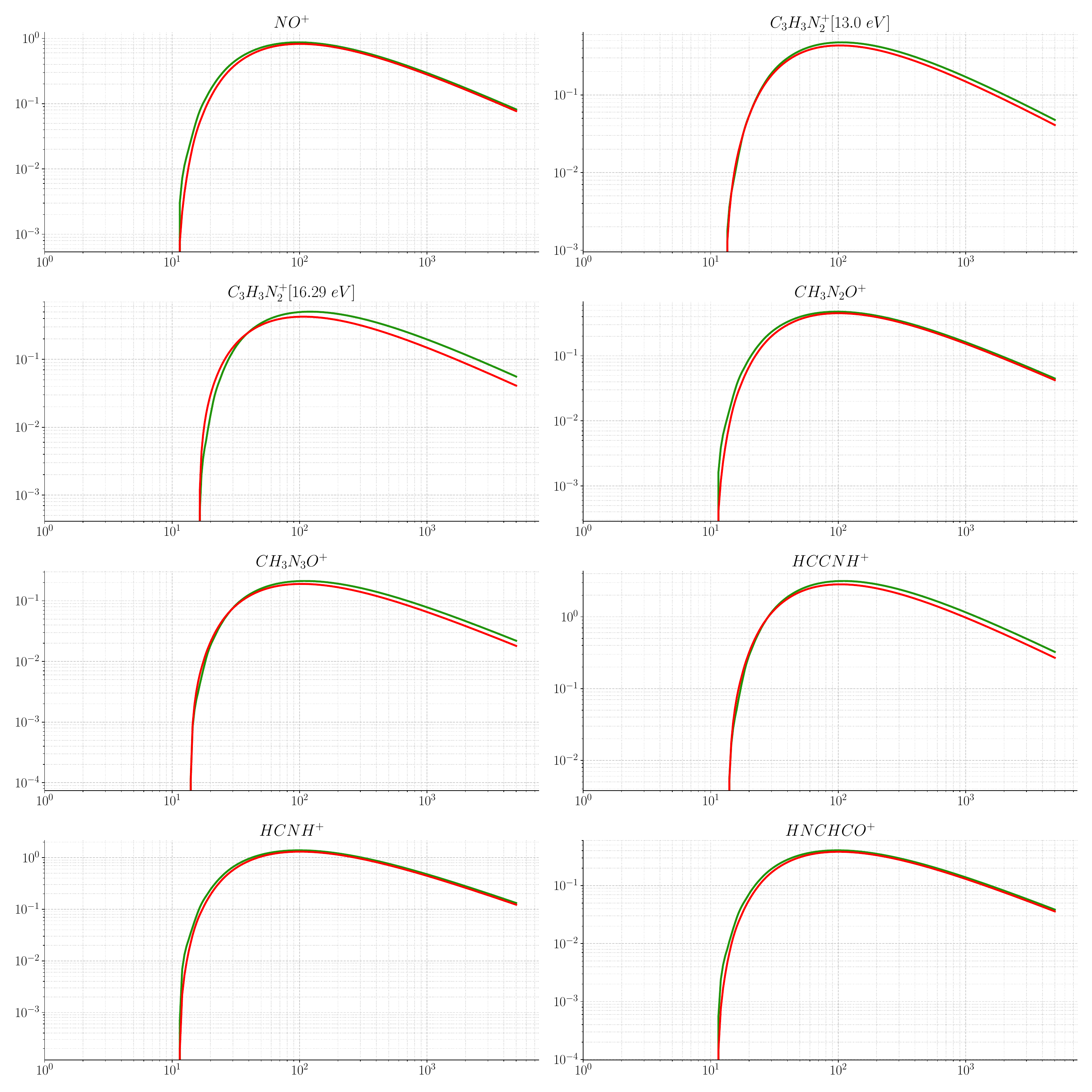}};
    \node[below, yshift=-0.5ex] at (current bounding box.south) {Energy (eV)};

    \node[left, rotate=90, xshift=9.5ex] at (current bounding box.west) {PICS $(10^{-16}cm^{2})$};
  \end{tikzpicture}
    \caption{}
    \label{fig:2NIMI-pics}
\end{subfigure}
\hfill
\begin{subfigure}[b]{0.45\textwidth}
    \centering
    \includegraphics[width=\textwidth]{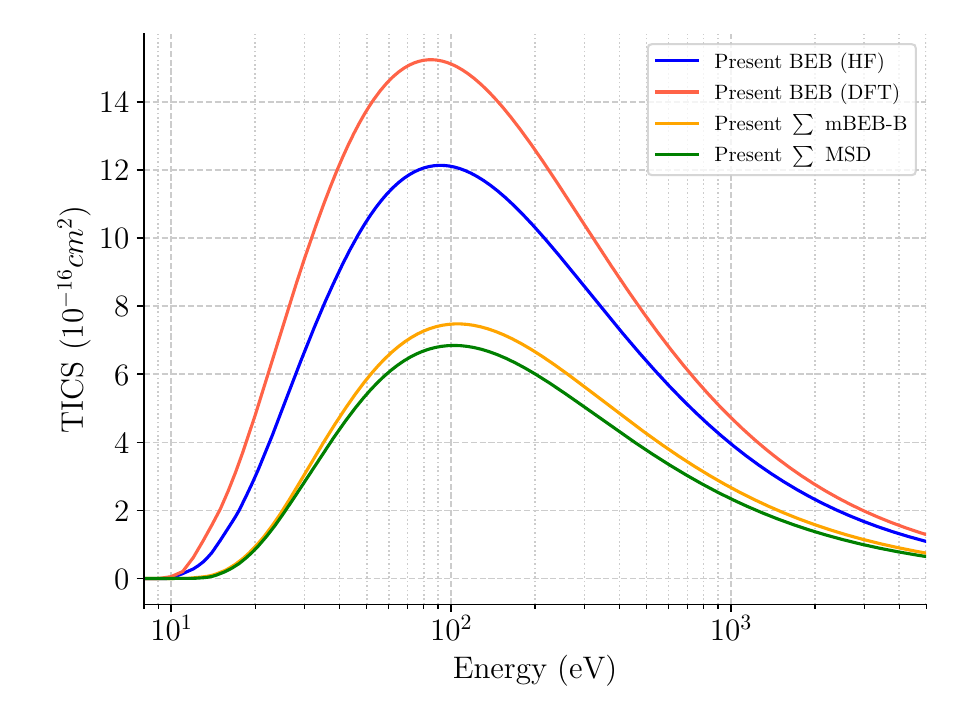}
    \caption{}    
    \label{fig:2NIMI-TICS}
    \end{subfigure}
    \caption{a) Comparison of PICS calculated using various methods for 2-nitroimidazole. The green line represents the PICS calculated using the mBEB-B method, and the red line represents the PICS calculated using the MSD method. b) Comparison of TICS for 2-nitroimidazole, the blue line indicates the present BEB data {for the orbital parameters obtained using the HF method, the peach colored line represents the present BEB data for the orbital parameters obtained using DFT method,} the green line shows the sum of PICS of MSD, and the orange line shows the sum of the PICS of the mBEB-B.}
\end{figure*}

The PICS of 2NI is shown in Figure \ref{fig:2NIMI-pics}, which contains the PICS of 8 fragments. The smallest fragment is $HCNH\p$ of m/z of 28 which has an AE of 11.44 eV and the biggest fragment is $C_3H_3N_3O\p$ having the m/z of 97. The $C_3H_3N_2\p$ cation has two AEs.
Like the PICS of IMI a similar trend is seen here, both the MSD PICS and the mBEB-B PICS show a good agreement from the ionization threshold to 100 eV, after which the PICS of the MSD method falls short on comparison with the mBEB-B PICS, but their magnitudes remain similar. 

The TICS of 2NI is shown in Figure \ref{fig:2NIMI-TICS}, with the BEB model compared with the TICS obtained after summing the cross sections of all the available fragments in the MSD and mBEB-B methods. The TICS do not match very well with the BEB TICS, as we don't have information on the AEs of all the fragments that appeared in the EIMS. All the details of the parameters required to compute PICS are given in Table \ref{tab:2Nimi_table}, which contains the theoretical BRs, the experimental BRs, the scaling factors, and the structural factors along with the maximum CS.
\begin{table*}
\begin{ruledtabular}
\caption{Absolute values of branching ratio $(\Gamma)$, structural factor $(\chi_i)$, scaling factor $(\Upsilon_i)$ and maximum PICS calculated using MSD and mBEB-B method compared for imidazole.}   
\centering
    \begin{tabular}{llllllllll}
         $m/z$&  Cations &  ${AE}$ (eV) \cite{meissner2019electron}&  $\Gamma_i^{H}$&  $\Gamma_i^{Theo} $&  $\Gamma_i$ \cite{meissner2019electron}&  $\chi_i$&  $\Upsilon_i$&  $\sigma_{max}^{MSD}$& $\sigma_{max}^{mBEB-B}$\\ \hline
         28&  $HCNH$&  11.74&  0.125833&  0.700872&  0.112307&  0.892509&  0.160239&  1.043251& 1.128965\\
         40&  $HCCNH$&  14.96&  0.060814&  0.498982&  0.246460&  4.052667&  0.493925&  2.240973& 2.592455\\
         40&  $HCNCH$&  14.96&  0.060814&  0.498982&  0.246460&  4.052667&  0.493925&  2.240973& 2.592455\\
         41&  $CH_2NCH$&  11.68&  0.127782&  0.705604&  0.025329&  0.198219&  0.035897&  0.235377& 0.254443\\
         41&  $CH_2CNH$&  11.68&  0.127782&  0.705604&  0.025329&  0.198219&  0.035897&  0.235377& 0.254443\\
         41&  $CHCHNH$&  14.06&  0.073256&  0.546657&  0.025329&  0.345758&  0.046334&  0.231722& 0.262657\\
         67&  $IMI-H$&  8.76&  0.302892&  0.688466&  0.007728&  0.025516&  0.054560&  0.073110& 0.378300 
    \end{tabular}
    \label{tab:imidazole-table}
\end{ruledtabular}
\end{table*}

\begin{table*}
    \centering
    \begin{ruledtabular}
\caption{Absolute values of branching ratio $(\Gamma)$ structural factor $(\chi_i)$, scaling factor $(\Upsilon)$ and maximum PICS calculated using MSD and mBEB-B method compared for 2-nitroimidazole.}  
    \begin{tabular}{cccccccccc}
         $m/z$&  Cations &  ${AE}$ (eV) \cite{meissner2019electron} &  $\Gamma_i^{H}$&  $\Gamma_i^{Theo}$&  $\Gamma_i$&  $\chi_i$&  $\Upsilon_i$&  $\sigma_{max}^{MSD}$& $\sigma_{max}^{mBEB-B}$\\ \hline 
         28&  $HCNH$&  11.44&  0.124877&  0.836333&  0.112307&  0.134285&  0.134285&  1.306890& 1.388680\\
         30&  $NO$&  11.11&  0.136339&  0.867858&  0.070544&  0.081285&  0.081285&  0.822306& 0.868480\\
         40&  $HCCNH$&  13.74&  0.072078&  0.654551&  0.246460&  0.376533&  0.376533&  2.832763& 3.155850\\
         56&  $HNCHCO$&  11.34&  0.128210&  0.845716&  0.032927&  0.038934&  0.038934&  0.383366& 0.406602\\
         67&  $C_{3}H_{3}N_{2}$&  13.00&  0.085100&  0.508922&  0.037563&  0.073808&  0.073808&  0.433492& 0.475207\\
         67&  $C_{3}H_{3}N_{2}$&  16.29&  0.043251&  0.508922&  0.037563&  0.073808&  0.073808&  0.425538& 0.504195\\
         83&  $C_{3}H_{3}N_{2}O$&  11.12&  0.135971&  0.866878&  0.038764&  0.044717&  0.044717&  0.451837& 0.477295\\
         97&  $C_{3}H_{3}N_{3}O$&  13.92&  0.069318&  0.642637&  0.016542&  0.025741&  0.025741&  0.189945& 0.212466 
    \end{tabular}
    \label{tab:2Nimi_table}
    \end{ruledtabular}
\end{table*}

\section{\label{sec:level5}Conclusion}

In the present work, we have theoretically computed the partial and total ionization cross sections of R-carvone, 2-butanol, imidazole, and 2-nitroimidazole using the MSD and mBEB-B methods in combination with the BEB model. We obtained a fairly good agreement of our theoretical PICS data of various fragments with the experimental PICS for R-carvone and 2-butanol. Also, the TICS which is obtained as the sum of all the PICS of the fragments in the MSD and mBEB-B method gives good agreement with TICS computed using the BEB model. The TICS obtained from the HF orbital parameters shows reasonable agreement with the experimental results for R-carvone and 2-butanol, whereas our DFT ($\omega B97X-D$/aVTZ) for R-carvone shows excellent agreement with the BEB data of Amorim \etal\cite{amorim2021electron} For imidazole and 2-nitroimidazole, we could not find any experimental PICS study and hence are computed in the present study for the first time. Hence, more investigation both theoretical and experimental is needed for these molecules. The appearance energies and relative abundances of the fragments were the key parameters for the determination of the PICS, which were all taken from the experimental papers for our calculations. All other parameters such as the experimental and theoretical BRs, scaling factors, and structural factors employed for the determination of PICS were computed and are presented in Tables. \ref{tab:r-carvone}, \ref{tab:2-butanol}, \ref{tab:imidazole-table} and \ref{tab:2Nimi_table} respectively. 
The TICS data obtained after summing all the PICS of the fragments in the MSD and mBEB-B method underestimates the TICS data computed using the BEB model for imidazole and 2-nitroimidazole. The discrepancy is due to the lack of complete information on the fragmentation of the parent molecule, where only the major dissociation fragments are given for imidazole and 2-nitroimidazole.  

\section*{Supplementary Material}

The supplementary material contains the optimized geometries of the target molecule, PICS of all the fragments of the parent molecules and the TICS of the molecules studied.

\begin{acknowledgments}
S.S acknowledges Vellore Institute of Technology (VIT) for providing a research fellowship; D.G acknowledges the Science and Engineering Research Board (SERB), Department of Science and Technology (DST), Government of India (Grant No. SRG/2022/000394) for providing financial support and computational facility.
\end{acknowledgments}

\section*{Data Availability Statement}

{The data that support the findings of this study are available in the form of supplementary material.}

\section*{Author Contributions}
\textbf{Suriyaprasanth Shanmugasundaram:} Conceptualization (lead), Data curation (lead), Investigation (lead),  Methodology (Supporting), Resources (Supporting), Validation (lead), Writing – original draft (lead), Writing – review \& editing (equal); \textbf{Rounak Agrawal:} Data curation (supporting), Methodology (Supporting), Investigation (Supporting); \textbf{Dhanoj Gupta:} Conceptualization (equal), Data curation (equal), Investigation (equal),  Methodology (lead), Resources (lead), Validation (equal), Writing – original draft (equal), Writing – review \& editing (lead), Supervision (lead).


\bibliography{main}

\end{document}